# A Framework for Facilitating Self-Regulation in Responsive Open Learning Environments

Alexander Nussbaumer (1), Milos Kravcik (2), Dominik Renzel (2), Ralf Klamma (2),
Marcel Berthold (1), and Dietrich Albert (1,3)

(1) Knowledge Technologies Institute, Graz University of Technology, Austria
{alexander.nussbaumer, dietrich.albert}@tugraz.at
(2) ACIS Group, Informatik 5, RWTH Aachen University, Germany
{kravcik, renzel, klamma}@dbis.rwth-aachen.de
(3) Department of Psychology, University of Graz, Austria
dietrich.albert@uni-graz.at

## Abstract

Studies have shown that the application of Self-Regulated Learning (SRL) increases the effectiveness of education. However, this is quite challenging to be facilitated with learning technologies like Learning Management Systems (LMS) that lack an individualised approach as well as a right balance between the learner's freedom and guidance. Personalisation and adaptive technologies have a high potential to support SRL in Personal Learning Environments (PLE), which enable customisation and guidance of various strengths and at various levels with SRL widgets. The main contribution of our paper is a framework that integrates guidance and reflection support for SRL in PLEs. Therefore, we have elaborated an operational SRL model. On that basis we have implemented a system with a learner model, SRL widgets, monitoring and analytic tools, as well as recommendation functionalities. We present concrete examples from both informal and formal learning settings. Moreover, we present analytic results from our SRL system – lab experiments and a public installation. With such a complex setting we are coming close to the realisation of Responsive Open Learning Environments (ROLE).

## 1 Introduction

Although nowadays a major goal of education should be to prepare people for lifelong learning, few students become intentional, independent, self-directed learners. Nilson (2013) considers that self-regulation includes the monitoring and managing of one's cognitive processes as well as the awareness of and control over one's emotions, motivations, behaviour, and environment as related to learning. She reminds that research has provided evidence that self-regulated learning enhances student performance in courses, the amount and depth of student thinking, students' conscious focus on their learning, as well as the development of reflective and responsible professionalism.

*Self-Regulated Learning (SRL)* has become a very active research field (Zimmermann 2002, Cassidy 2011, Sabourin et al. 2013). The focus of educational research has been on the cyclic phase structure of the SRL process as well as its meta-cognitive, motivational and behavioural aspects. We are taking this background into account, but have a *Technology Enhanced Learning (TEL)* perspective on SRL. Our main interest is the use of *Personal Learning Environments (PLE)* to facilitate SRL processes. PLEs (Olivier and Liber 2001, van Harmelen 2006) to a large extent consist of Web 2.0 resources and tools (Klamma et al. 2007). It is assumed that learners define their own learning goals and manage their learning contents or processes with a high degree of autonomy (Downes 2007). Therefore a main aim is to create PLEs that effectively support SRL, meaning that they enable and help learners to apply SRL activities. This is effective, because the repetitive application of SRL activities leads to an increase of the SRL abilities.

We argue that SRL processes in reality are embedded also in many learning settings. Therefore in our work we consider several test beds ranging from formal to informal ones. SRL happens even within formal learning processes in educational institutions. We also argue that in many learning situations the learner is not ready to learn autonomously. Therefore, we claim that current research is not taking into account the complexity coming from differences in technology and learning culture as well as the real needs of real learners. This summarizes the *openness* of the approach.

The concept of *adaptivity* is traditionally based on the domain, the user and the context model. But there are often problems with the identification of user preferences. One reason is that user preferences can change very quickly and these changes may be difficult to recognize. Another issue is that people often do not consciously know their preferences (Dijksterhuis 2006) and their actions alter their preferences – when they select something, they will value it more (Sharot et al 2010). Unknown preferences may be revealed by facing choices. Although uncertainty is perceived usually as unpleasant, it has also a positive side, which includes the option to change – the ability to switch from a course of action. Nature illustrates how optionality is a substitute for intelligence and the option is a substitute for knowledge (Taleb 2012). Options and their implementations provide the mechanism to facilitate various degrees of guidance and freedom. Based on the nature of PLEs and the existence of SRL in various learning settings, we do not consider the traditional concept of adaptivity. Instead we pursue a libertarian paternalism approach (Thaler & Sunstein 2008), offering the right mix of adaptivity and recommendations of available options. This is our understanding of *responsiveness*.

Openness and responsiveness have been underestimated in research so far and lead to a new class of personal learning environments that we call *Responsive Open Learning Environments (ROLE)*. We make the following contributions to research:
- We define a conceptual model of the self-regulated learning and relate it to the responsiveness and openness of the learning environment.
- We present a technical framework for supporting SRL in ROLE, allowing us not only to support self-regulated learners, but also to serve as a research platform for SRL analytics.
- We present several authentic learning settings and demonstrate the SRL support in a conceptual and technical way; with respect to the learning settings, we give detailed and comprehensive evaluations of the SRL processes.

The rest of the paper is organised as follows. In Section 2 we summarize the SRL research background. We then present our conceptual model in Section 3, including two authentic learning scenarios. In Section 4 we introduce our technical infrastructure in a condensed fashion. The evaluation in Section 5 shows detailed results from a controlled study, a summary of the outcomes from our test beds, and SRL analytics results from learning processes performed in our public installation, the ROLE Sandbox. Finally, we conclude and outline future work in Section 6.

## 2 Theoretical Background of SRL

In the recent decades much research has been conducted in the field of SRL, which overlaps with different disciplines, including pedagogy, psychology, neuroscience, and technology-enhanced learning. This section provides an overview of the theoretical background of SRL. It clarifies why it is important to facilitate and train meta-cognitive skills (like SRL) and it highlights which mechanisms can be used for this purpose.

A revolutionary discovery in brain science was neuroplasticity (Doidge, 2008) – our thoughts can change the structure and function of our brains. It means that our experience and learning can switch our genes on and off, altering our brain anatomy and our behaviour. This enables endless adaptability of the human brain and also confirms that SRL skills are trainable. Neuroscientists distinguish between conscious, declarative memory about facts and events, versus unconscious, non-declarative memory abilities, like skills and habit learning (Kandel & Squire, 2000). The different brain substrates forming non-declarative memory are evolutionary ancient, which leads to habits that influence our behaviour. Habits can be defined as *"response dispositions that are activated automatically by the context cues that co-occurred with responses during past performance"* (Neal et al., 2006). They help save brain effort and drive much of everyday actions with minimal conscious control. They are difficult to change, which has crucial consequences for behaviour and self-regulation of humans. Habit memory involves slowly acquired associations between stimuli and responses. Humans possess a robust capacity for gradual trial-and-error (habit) learning that can operate without awareness for what is learned (Bayley et al., 2005).

Human mental processes can be described from the psychological perspective by the metaphor of two agents, called System 1 and System 2 (Kahneman, 2011). System 1 is fast, impulsive and intuitive, but prone to cognitive biases. If System 1 is involved in cognitive processes, the conclusion comes first, and the arguments follow. System 2 is capable of reasoning, cautious, but in some cases "lazy". Self-control and cognitive effort are forms of mental work, and System 2 controls thoughts and behaviours. This means that System 2 monitors and controls actions "suggested" by System 1. Various activities compete for the limited resources of System 2, so self-control and deliberate thought draw on the same limited budget of effort. Correlations were found between self-imposed delay-of-gratification in preschool versus cognitive competence and ability to cope with stress in adolescence (Shoda et al, 1990). This indicates that self-regulatory competences can be recognized very early and have a high impact on successful learning during the lifetime. While according to neuroplasticity new skills can be acquired at all times, these results also suggest the necessity of early cultivation of self-regulatory skills.

Self-regulation is crucial for the development of lifelong learning skills. According to educational psychologists, SRL is guided by meta-cognition, strategic action, and motivation to learn (Zimmermann, 1990). In this context students are proactive with respect to their learning (Zimmermann, 2002). Research shows that self-regulatory skills can be trained and can increase students' motivation and achievement (Schunk & Zimmerman, 1998). Our aim is to facilitate usage and training of SRL skills by employing responsive open learning environments in the learning process.

Regarding learning performance, there is evidence that students with intrinsic motivation, initiative, and personal responsibility achieve more academic success (Zimmermann & Martinez-Pons, 1990). Studies also indicate that in order to improve academic achievements, all three dimensions of SRL in students must be developed: the meta-cognitive, the motivational, and the behavioural ones (Zimmermann, 1990). Another interesting finding is that SRL can enable accelerated learning while maintaining long-term retention rates (Lovett et al., 2008). In a meta-analysis of 800 meta-studies (Hattie, 2009), it has been shown that applying meta-cognitive learning strategies significantly contributes to learning success. These results provide clear evidence that meta-cognitive skills and in particular SRL abilities belong to the key competences of a successful learner, especially in the context of lifelong learning.

Azevedo and his team demonstrated the effectiveness of SRL training (short-term – 30 min) in facilitating students' learning. They concluded that learners who received this training could effectively deploy the SRL processes leading to significant shifts in their mental models (Azevedo & Cromley 2004). In their empirical research this team investigated how students regulate their learning with advanced learning technologies and provided guidelines for supporting SRL with their hypermedia learning environment MetaTutor (Azevedo et al. 2010). Their results indicated that SRL training was effective (Azevedo et al. 2009). As students with metacognitive (e.g. SRL) skills are more successful learners, the aim of a recent study (Sabourin et al., 2013) was to identify these skills early enough. This knowledge can be used later on for suitable training or adaptive scaffolding. For their predictions the authors used highly structured game-based learning environments. This provides motivation to repeat similar experiments also in more open-ended exploratory environments. It is important to measure the impact of adaptive scaffolding both in terms of learning performance and user engagement.

Various conceptual models of SRL have been proposed. The layered model (Boekaerts, 1999) represents *regulation of the self* (choice of goals and resources), *regulation of the learning process* (use of metacognitive skills), and *regulation of information processing modes* (choice of cognitive strategies). The model has been informed by three schools of thought – theories of the self, research on metacognition, and learning style research. Based on them, learning style, academic control beliefs, and student self-evaluation have been identified as critical factors in our understanding of student academic achievement and key processes of SRL (Cassidy, 2011). In order to measure SRL skills through social interactions, Biswas et al. (2010) developed a metacognitive model that interprets students' activity sequences as aggregate patterns of their learning behaviours. They link these patterns to students' use of self-regulated learning strategies. These models informed the development of our pedagogical model, which will be introduced in Section 3.

An important question relates to the design of virtual and intelligent learning environments for SRL. According to one case study (Whipp & Chiarelli, 2004) relevant motivational influences on SRL strategy use (self-efficacy, goal orientation, interest, and attributions) were shaped largely by student successes in managing the technical and social environment. In addition, important environmental influences on SRL strategy use included instructor support, peer support, and course design. Researchers have identified the importance of cues (like nudges or recommendations) in the learning environment that impede and facilitate self-regulation (Boekaerts & Cascallar, 2006). Reports have been published (Duckworth et al., 2009; Meyer et al., 2008) that provide guidance on the implementation of SRL in school education. These mention that SRL improves with practice, requires an 'enabling environment', and can be supported by ICT tools. Moreover, any interventions to promote SRL are likely to be long-term. One of the existing tools leverages ontologies to support self-regulation in organizational learning (Siadaty et al., 2011). A review of recent advances in learner and skill modelling in intelligent learning environments (Desmarais & Baker, 2012) mentions modelling meta-cognition and SRL among the main advances.

Cognitive biases and other limitations of the human mind play an important role in the decision making process. Choice architecture describes the way alternative items are presented to the chooser. It is pervasive, unavoidable, and can have massive effects on people's behaviour. *Libertarian paternalism* approach has been proposed (Thaler & Sunstein, 2008) to preserve liberty and to influence choices in a way that will make choosers better off, as judged by them. This can be realized via suitable nudges, which should alert people's behaviour in a predictable way and at the same time should be easy and cheap to avoid. The golden rule of libertarian paternalism states: offer nudges that are most likely to help and least likely to inflict harm.

The principles of libertarian paternalism suggest that a suitable way to support SRL is by means of flexible and adaptable learning environments that provide enough freedom as well as context-dependent recommendations on demand. This should lead to the right balance between the freedom of choice, which is crucial for motivation, and appropriate guidance, in order to help making the learning process effective and efficient. This balance depends on the context, including the learner, the subject domain, and the current constraints. From our perspective this means the freedom to organize and control one's own learning process, design the PLE, and choose learning resources, as well as an opportunity to receive suitable, context-dependent guidance in the form of recommendations, together with the opportunity to select or avoid the provided offers.

# 3 ROLE Approach and Scenarios

The vision of the ROLE approach is to empower the learner to create her own learning environment that is suitable for self-regulated learning. Considering that the ROLE platform integrates a great number of learning resources (tools, services, content, and peers), we designed a psycho-pedagogical framework to support the individual composition of learning services and their usage. Our framework shows how SRL support can be provided and is broader than it has been implemented so far. Generally, this support is based on the provision of a PLE where learners are motivated and stimulated to apply SRL activities. This section presents also concrete examples and usage scenarios that demonstrate how the ROLE platform has been used.

## 3.1 Psycho-Pedagogical Approach

Components of SRL are cognition, meta-cognition, motivation, affect, and volition (Kitsantas, 2002). Six key processes that are essential for self-regulated learning are listed by Dabbagh & Kitsantas (2004). These are *goal setting*, *self-monitoring*, *self-evaluation*, *task strategies*, *help seeking*, and *time management*. A cyclic approach to model SRL has been given by Zimmerman (2002), where SRL is seen as a process of meta-cognitive activities consisting of three phases, namely the forethought phase (e.g. goal setting or planning), the performance phase (e.g. self-observation processes), and the self-reflection phase. According to this model learning performance and behaviour consist of both cognitive activities and meta-cognitive activities for controlling the learning process. Aviram et al. (2008) have extended this model towards a Self-Regulated Personalised Learning (SRPL) approach by adding a self-profile so that learners can indicate their own preferences.

Based on these findings Fruhmann et al. (2010) have developed a learning process model that takes into account the requirements to create and use responsive open learning environments that support SRL. It is adjusted to ROLE in two ways. First, an additional phase has been introduced that is primarily related to the creation of the learner's own learning environment (preparation phase). Second, the phases are composed not only on the meta-cognitive level, but also on the cognitive one. The reason for these adaptations stems from the consideration that all learning activities relevant to the creation and usage of a personal learning environment should be included in this model. So this SRL process model results from the collection of learning activities performed by the learner using the ROLE platform. A categorisation of these learning activities resulted in the four learning phases, which are *planning*, *preparing*, *learning*, and *reflecting*. Figure 3.1 depicts this model with the four phases and related learning strategies.

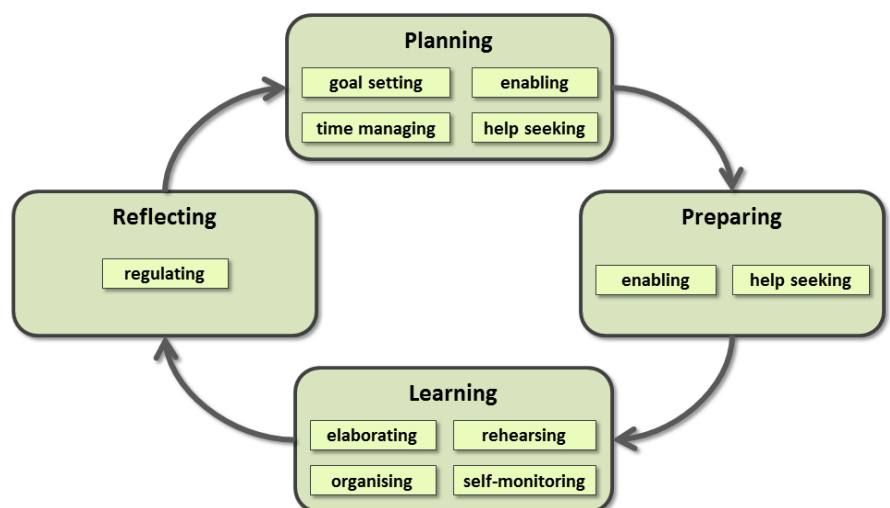

Figure 3.1. The self-regulated learning process model consists of four phases and related learning strategies.

Following the ideas presented in (Mandl & Friedrich, 2006), learning strategies and learning techniques are used to describe learning activities and are associated with the cyclic phases of the learning process model. In this way we describe the phases with clearly defined activities. Learning strategies describe the conceptual background, when self-regulatory activities are applied. More precisely, learners apply learning techniques (such as brainstorming, goal setting, or note taking) which are categorised according to learning strategies. A learning strategy is rather an umbrella term to classify learning techniques and various authors provide different presentations of how they relate to each other. Learning techniques in turn are often defined as sophisticated methods for implementing learning activities. Learning strategies can be understood as the *what* (what I want to do) and learning techniques as the *how* (how I perform the learning activity). Typically learning tools or widgets are used for applying learning techniques. Therefore, our model also includes associations between learning techniques and learning tools.

Based on the analysis of SRL processes described by Dabbagh & Kitsantas (2004) and the description of learning strategies and techniques by Mandl & Friedrich (2006) we identified nine SRL strategies and structured them in three groups: (1) *cognitive strategies*, (2) *meta-cognitive strategies*, and (3) *resource management strategies*. The group of cognitive strategies includes organisation, elaboration, and rehearsal strategies (regarding learning topics). The group of meta-cognitive strategies includes goal setting, self-monitoring, and regulation strategies targeting the control of one's own learning process. The group of resource management strategies includes time management, help-seeking, and enabling (or environment preparation) strategies employed by the learners to manage their learning resources and environment. For each of these nine learning strategies a variable number of learning techniques is assigned. For instance, elaboration can be done by applying techniques like paraphrasing, creating analogies, producing questions, note taking, brainstorming, or collaborative learning. In total 37 learning techniques have been identified and assigned to the nine aforementioned learning strategies. Furthermore, all these techniques are associated with learning tools from the ROLE Widget Store. Though we find this taxonomy as useful, it is still possible to extend or change it.

In addition to learning activities, competences are important and powerful constructs for describing and characterising learners. In our competence model we distinguish three types of competences, originating from the fact that learning happens on different competence levels, which are knowing a subject domain, being able to learn by oneself, and being able to use tools or widgets for learning purposes. Therefore, three types of competences are defined as *domain competences*, *SRL competences*, and *tool competences* (Fruhmann, 2010).

According to the European Qualifications Framework for Lifelong Learning (EQF, 2008), domain competences are defined by a pair of *domain concept* and *proficiency level*. The proficiency level defines how well a domain concept is known by the person. Concepts are defined as a pair of a specific concept and its context. The context is the vocabulary or taxonomy in which the concept is defined. For example, it can be a concept map, but also DBPedia, which semantically describes Wikipedia with all article titles as unique concepts. Tool competences are defined as a pair consisting of a *learning tool* and a *learning technique*. Since ROLE focuses on the individual composition of learning environments consisting of learning tools, the usage of tools for learning is addressed in this competence definition. In general, tool competence defines the ability of learners to learn (to attain domain knowledge) with a specific tool. The SRL competence is defined as a pair of SRL strategy and *EQF level*. This definition includes the strategy that is applied for self-regulated learning. The EQF level determines how well such a strategy can be applied. SRL and tool competences are regarded as independent from domain knowledge.

Supporting SRL is a major aim of the ROLE approach. Based on Efklides' (2009) proposal, we identified five key dimensions as important for supporting self-regulated learning: (1) *adaptability*, (2) *guidance and freedom*, (3) *meta-cognition and awareness*, (4) *collaboration* and (5) *motivation*. Adaptability targets the possibility to customise the learning environment. The learner should be empowered to design a PLE according to her needs and preferred learning techniques, content, tools, services, peers, and communities. The degree of guidance and freedom is an important factor in providing the optimal support for learners. While some learners will not need (or want) any support, others will require guidance for creating and using the learning environment in a self-regulated way, as it turned out in real-life experiences (see Section 5.2). Therefore, different levels and types of guidance are needed for learners. The common feature of all these types of guidance is that they are not forcing learners or prescribing what to do, but leave the decision for the learner. In our case guidance is realised by providing recommendations of SRL activities, widgets, and content.

Learners should also be supported in their meta-cognition, because this is specifically important for SRL. Therefore suggestions are made to apply learning activities on the meta-cognitive level and to use learning tools for their support. Collaborative learning comprises individually performed activities and also extra activities that are generated by interaction among peers (Dillenbourg, 1999). These collaborative activities trigger additional cognitive mechanisms and appear more frequently in collaborative learning situations than in individual learning. Motivation is crucial for effective learning. To stimulate intrinsic motivation there is a need for autonomy, competence, and relatedness (Deci & Ryan, 2000). In our approach, motivation is a consequence of the right balance between guidance and freedom (as explained in Section 2). Hence, motivation is an implicit feature of the overall framework. However, in some cases initial briefing is needed. Learners reacted positively to video material that simply and easily explained the core concepts (e.g. Nussbaumer et al., 2012c).

## 3.2 ROLE Framework

**Space concept and technical approach**

From the user's perspective the central piece of our learning environment is the ROLE Space, which provides the user interface for the learner. The ROLE Space is a container of widgets that usually consist of a front-end and a service in the background. Examples of such widgets are a search tool for learning material or a tool for planning learning activities. The user (learner or teacher) can select widgets from a repository (ROLE Widget Store) and add them to a personal environment. Our implementation provides several core features of the space concept. First, it provides a dynamic storage of user resources where learner-specific information can be stored, such as competences or goals. Second, it enables a communication infrastructure, which allows widgets to communicate with each other in the same space. Third, it offers a collaborative approach, enabling learners to communicate with other users in the same space. Fourth, a log data format has been specified and implemented. Putting together these features, a ROLE space can be defined as a bundle of widgets that includes storage for user information and widget data. The spaces can also be pre-configured and shared with others, so that the users of the same space can communicate with each other directly or indirectly over the widgets. To support learners, recommendations are made on different levels. In principle, learners are provided with recommendations in terms of resources (content artefacts, widgets, or peers) or learning activities. User information regarding competence, goals, learning history and preferences provide the basis for the recommendations. Feedback is given based on the data available in the user model. In order to acquire user model data, the learner can provide data directly, or a monitoring approach is applied. All these data are based on the ROLE ontology as described in the following. An overview of this framework is depicted in Figure 3.2.

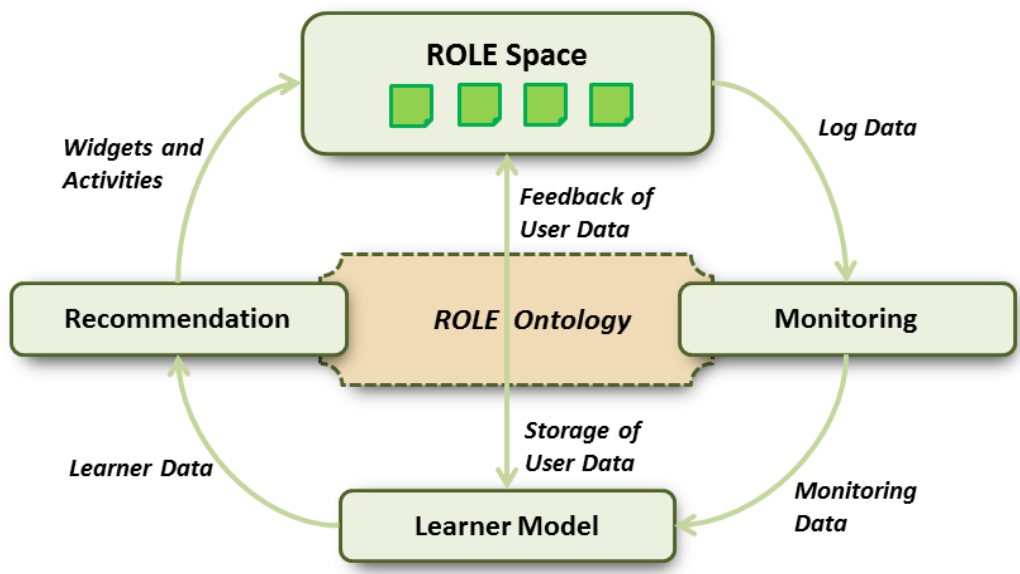

Figure 3.2: The conceptual technical approach to support SRL.

**Learner Model**

A recent review (Desmarais & Baker 2012) mentioned several important advancements in this area, including modelling learner motivation, emotional state, meta-cognition and SRL, as well as open and long-term learner modelling. Research shows that to support self-monitoring of long term goals people need meta-cognitive scaffolding (Tang & Kay 2013). A newer trend follows the idea of opening up these models to the user, in order to support the learners in their reflection learning process and to support teachers to better understand their students. A promising avenue for facilitating reflection and raising awareness is represented by Open Learner Models (OLM), where visualization, trust, and credibility play key roles (Bull & Kay, 2010). OLMs provide suitable interfaces for users to enable them to view and in some cases also to change their learner model. This information can be available also to others (peers and teachers), who can assist in the learning process of an individual. Examples of OLM visualisation techniques are overview-zoom-filter approaches including tree maps, tag clouds, or sunburst views (Bull & Kay, 2010; Mathews et al., 2012; Conejo et al., 2011).

In adaptive systems, user models are often designed as overlay models (Brusilovsky & Millan, 2007), where user information relies on and is described by conceptual information (e.g. concepts maps). Our learner model follows this scheme, but also relies on the activity model and its elements. This became necessary, because the information about the learner is not only given at domain level, but also at the level of cognitive and meta-cognitive activities. The learner model consists of four elements. First, the competence state describes the available competences of the learner taking into account that there are three different types of competences: (i) domain knowledge, (ii) ability to learn with tools, and (iii) ability to learn in a self-regulated manner. Second, the goals of a learner are also described using these competences. The learning goal is expressed in terms of the competences a learner wants to achieve. Third, the learning history is described by the learning activities (strategies and techniques) a learner has applied, the tools or widgets she has used, and the competences she has attained. In contrast to low level log data, this learning history provides more insight because it describes the history not just in terms of particular interactions, but also regarding learning activities and used resources. Fourth, pedagogical parameters describe individual preferences and properties of the learner. Examples include information about preferred tools, learning groups, or guidance mechanisms. An overview in the context of the other models is given in Figure 3.3.

The formal representation of the learner model is included in the ROLE Ontology in RDF/OWL format. Because it is only understandable when described with the related concepts from the other models already mentioned, we use the opportunity to describe the classes and the properties very briefly here. An overview of this ontology is depicted in Figure 3.3. The Competence class is the superclass of the three types of competences. It allows specifying title and description reusing these properties from the Dublin Core terms vocabulary (DC...). Furthermore, it provides two properties that allow for a generic definition of the competences. First, a topic property is used to relate a specific competence to a generic object. Second, the *proficiencyLevel* property is used to express the extent to which a user has a specific competence. The generic Level class is subclassed by the EQF class to indicate the competence level.

In order to relate a competence to a learner, a Person class is used that has the properties *acquiredCompetence*, *goalCompetence*, *uses*, *applies*, and *hasParameter*. In this way, competences are defined that a learner has already acquired and competences that a learner should attain, the tools a learner has used, the learning techniques she has applied, and the personal preferences. The *DomainCompetence* class allows for defining a domain competence. It is a subclass of the Competence class and inherits title, description, topic, and the EQF level specification. In this way, competence can be defined by using the topic property to assign the related domain concept and the *proficiencyLevel* property to define the EQF level. The *ToolCompetence* class enables the definition of a tool competence. It is a subclass of the Competence class and inherits the specification of the title and the description. Following the definition of tool competences, the topic property is used to relate tool functionalities. The technique property is used to relate a *LearningTechnique*. The SRLCompetence class allows the definition of SRL competences. It is a subclass of the Competence class and inherits the specification of the title and the description as well.

Following the definition of SRL competences, the topic property is used to relate to SRL strategies. The inherited *proficiencyLevel* property is used to relate to an EQF level. Since SRL competence can be related to different types of SRL strategies (cognitive, meta-cognitive, and resource management strategies), appropriate subclasses of the *SRLStrategy* class have been defined.

The learning activities (SRL phase, SRL Strategy, Learning Technique) are also modelled and accessible, but not described here. In total, it consists of 4 SRL phases, 9 SRL strategies, and 31 learning techniques.

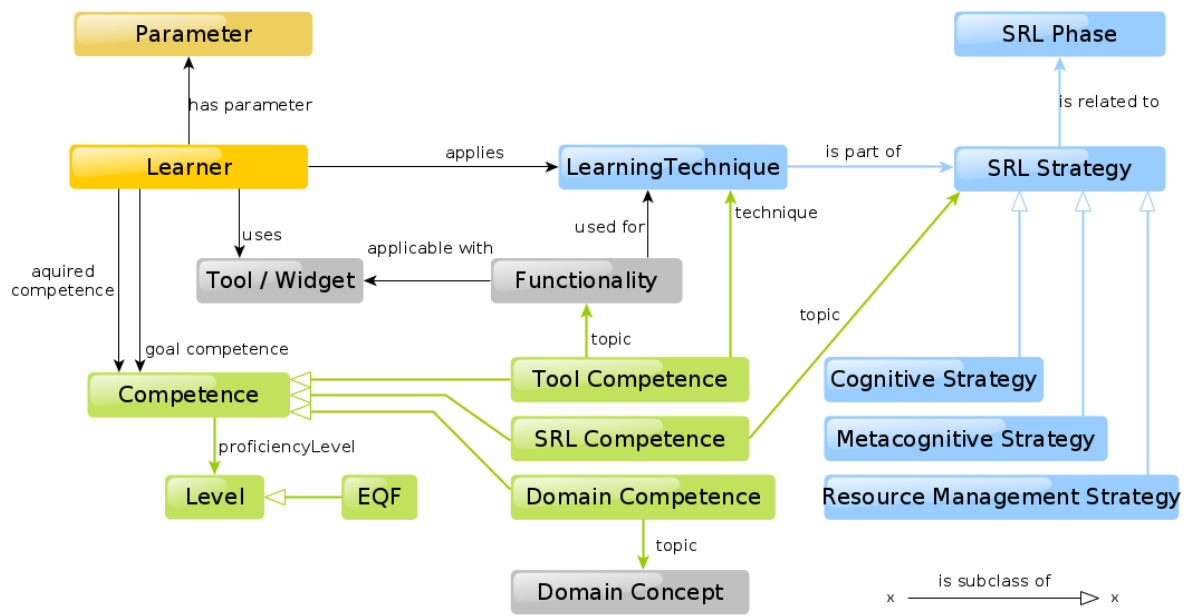

Figure 3.3: The ontology of the SRL models as part of the learner model.

**Personalisation and Recommendation**

Traditional adaptive systems adapt the content and system behaviour automatically to the learner by using the information from the user model, which is also called personalisation. Though this approach might be very effective for acquiring domain knowledge, it does not take into account learning aspects relevant for SRL. For example, planning includes the task of selecting learning resources considering the own goals. Therefore, the ROLE approach focuses on recommendations, which gives control and guidance to users at the same time. Like traditional adaptation, the recommendation approach is based on a user model and a recommendation strategy. In our approach, three types of resources can be recommended. These are widgets or learning tools, learning activities, and content or learning objects.

Widget recommendation helps the learner to create a PLE. Recommending widgets for the widget space is based on a template of SRL activities. The template is a set of SRL phases, strategies, and techniques supporting certain learning scenarios, such as language learning. When the learner clicks on an entity of the template, a set of widgets is recommended from which the learner can choose. A simple click on a recommended widget adds this widget to the current space. This recommendation is created by exploiting the relations between learning strategies, techniques, and widgets. The learner model information is used to rank the widgets. For example, if a learner has the goal to improve her reflection competence, then widgets are ranked higher if they are associated with learning techniques being part of the reflection strategy. This approach has been implemented as the Mashup Recommender widget and is described in detail in (Nussbaumer et al., 2012a).

Learning activity recommendation is a feature that guides the learner through the learning process. Learning phase, strategies, and techniques are recommended in a sequence and frequency that ensures that different types of learning activities are covered. Based on the applied activities, the learner model is updated, and new recommendations are provided. The learner can also choose to skip an activity or to select an activity and request further activities on a deeper level (e.g. learning techniques for a selected learning strategy). In combination with the widget recommendation approach, a learner can also get widgets for a selected learning technique. This approach provides guidance for learners with different SRL competence levels. Learners with higher SRL competency are able to skip recommendations in case the recommended information is already known or too detailed for them. Otherwise, the learner is guided by a step-by-step approach helping to cope with a problem. Unlike direct instructions, the learners have a free choice of recommended learning activities they want to perform. More information can be found in (Dahn et al., 2013).

Content recommendation is based on concepts included in domain competences a learner has or wants to acquire. Such competences are stored in the learner model. Since domain competences include concept definitions, this information is used to search learning object repositories. The result of such a query is presented as a recommended learning object. Details are described in (Moedtritscher, 2012).

**Mashup Design**

We have developed guidelines for creating pedagogical mashup designs. The most important ones are listed here. The validity of our guidelines has been partially confirmed by the studies reported in Section 5. The design of a PLE is an important concern, because learning is affected by the included components. Compiling different widgets or widget bundles for a PLE relates to different psycho-pedagogical aspects and educational strategies. We identified different educational components that might influence learning: the learning goal to be achieved, the learning strategies and techniques applied, the widgets used for applying them, and the peers taking part in collaborative learning.

There is a connection between learning techniques and widgets. Certain widgets can be used to apply particular learning techniques. According to the idea of SRL, learners can apply different learning strategies to control their own learning process. Therefore, a mashup design should contain widgets that support various learning strategies, for example one of each phase of the SRL process model. This can be realized in a PLE with a widget for goal setting, a widget for content searching, a widget for organizing and self-monitoring, and a widget where the learning process is visualized (reflection). Another guideline takes into account the tool competences. As described above, a tool competence is defined as the competence to be able to successfully learn with a tool or widget. Accordingly, a mashup design should contain widgets that are suitable for a specific learner. The next guideline is related to the number of widgets in a learning environment, which also plays an important role. Too many widgets and their different functionalities might overwhelm the learners at first, especially if they are not intrinsically motivated to perform the learning task (Ulrich, Shen & Gillet, 2010). A good number of widgets also depends on the individual skills and preferences of the learner. Furthermore, the subject domain addressed by the PLE is highly relevant. Widgets and resources are needed that support the achievement of the learning goals. In order to support the SRL process, recommender tools for learning techniques or even concrete activities might be useful. Finally, also peers are relevant in a collaborative PLE. Peers can help by playing the role of a tutor or they might be helpful for just discussing learning topics and problems, which in turn stimulates reflection.

## 3.3 Usage Scenario 1: SRL Text Reader Bundle

The SRL Text Reader Bundle (Figure 3.4) is a predefined widget bundle that supports SRL by providing feedback on the SRL activities. The bundle captures certain SRL activities and displays them in a graphical way to make the learner aware about her behaviour. The main widget is the *Text Reader* where learners read and annotate texts. Such texts and related concepts are defined in the domain model. The *Self-Evaluation* widget allows to relate the assigned tags with concepts from the domain model and to determine the proficiency level for each concept. In this way, the learner evaluates herself regarding her own domain competences. The *Modified Binocs* widget allows searching additional resources for the domain according to related tags and concepts. All performed activities are recorded and stored in the learner model. The *Self-Reflection* widget follows our Open Learner Model approach and gives feedback to the learner about her learning process. In addition, this widget displays the texts that have been annotated and the concepts that have been used for self-evaluation. Guidance for the whole SRL process is provided by this set of four widgets.

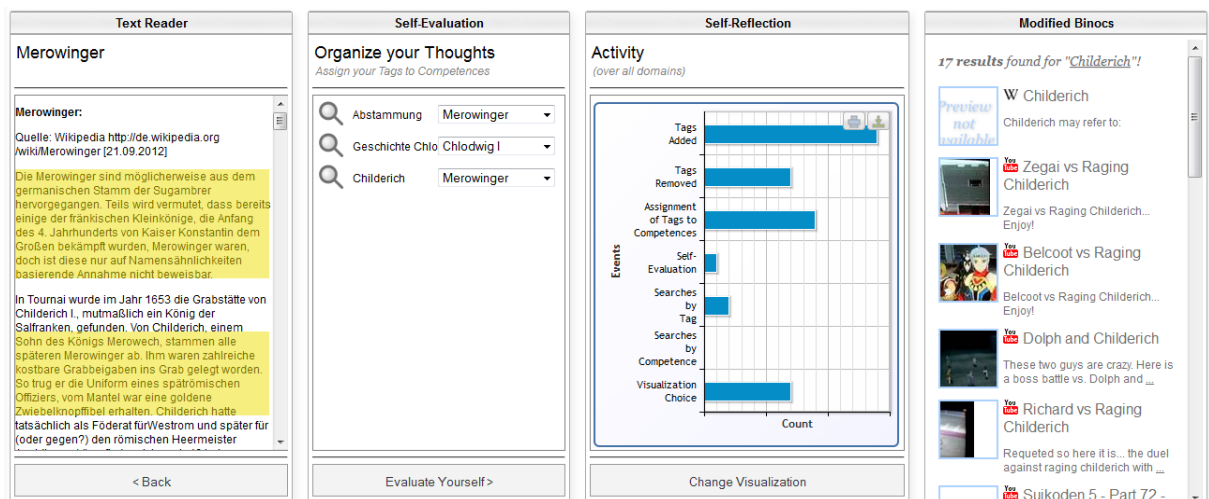

Figure 3.4: The SRL Text Reader Bundle. This bundle consists of four widgets that together support the application of different cognitive and meta-cognitive learning strategies.

*Scenario Description: A teacher prepares a lecture on a history topic for his students. He writes several texts, structures the topic with the help of a concept map, and assigns the*

*concepts to the texts. Then he creates a domain model consisting of texts and concepts and includes it in the backend service of the SRL Text Reader. Finally, he instantiates the SRL Text Reader Bundle in a space and shares the space's URL with his students. A student receiving the URL navigates to the space and sees the four widgets. She starts with the Text Reader widget reading the prepared texts. While reading through the texts, she marks and tags several paragraphs. On a cognitive level, marking and tagging means organising and structuring information. From time to time, the student switches to the Self-Evaluation widget where both the concepts of the teacher and the tags of the student are displayed. She assigns the tags to the concepts and self-evaluates how well she has understood the concepts. By clicking on a tag or a concept in the Self-Evaluation widget, a search for additional learning materials is triggered automatically in the Modified Binocs widget. Activities in all these activities are recorded and sent to the learner model. The Self-Reflection widget displays the number of activities, thus giving an immediate overview of the student's behaviour. For example, it shows if the student has only read texts, but did not tag them and did not perform self-evaluation.*

From the learner model perspective, different types of information are saved for further usage. The activities performed by the learner are saved using the activity schema (outlined in the ontology definition in Section 3.2). The concepts used for self-evaluation are coming from a domain model stored together with the proficiency level as domain competence. The tags related to certain texts are saved as generic information. All this information is used for keeping the user data persistent and for doing analysis visualised in the Self-Reflection widget.

A modification of this bundle was made by adding the *SRL Monitor* widget (preliminary outline is available in Nussbaumer et al., 2012b), which is a generalisation of the *Self-Reflection* widget and provides support in developing self-awareness about the learning activities performed. The goal is not only to monitor and visualize the observable actions (as saved in the log data), but also to monitor the cognitive and meta-cognitive activities that are not directly measurable. To this end, the measurable actions are mapped to cognitive and metacognitive learning activities from the ontology. To be precise, the key actions extracted from the log data analysis (based on an algorithm that clusters the log data) are mapped to elements of the learning ontology. The mapping is partially done by the learner herself, but also supported by an algorithm that takes into account the previous manual assignments. The goal is to make learners aware of their cognitive and metacognitive learning activities.

The screenshot displayed in Figure 3.5 shows two views of the SRL Monitor. In the first view, the SRL Monitor displays the learner's captured log data in a sequence. Then, the learner can select which learning technique she actually applied. Based on these selections, reasoning is done regarding the applied learning strategies. Since there are only 9 learning strategies, a comprehensive overview of the learner's behaviour can be given. This overview is graphically shown in the second view. In this way, learners get feedback about their learning behaviour and might rethink their learning process if some learning strategies never appear in the graphical profile.

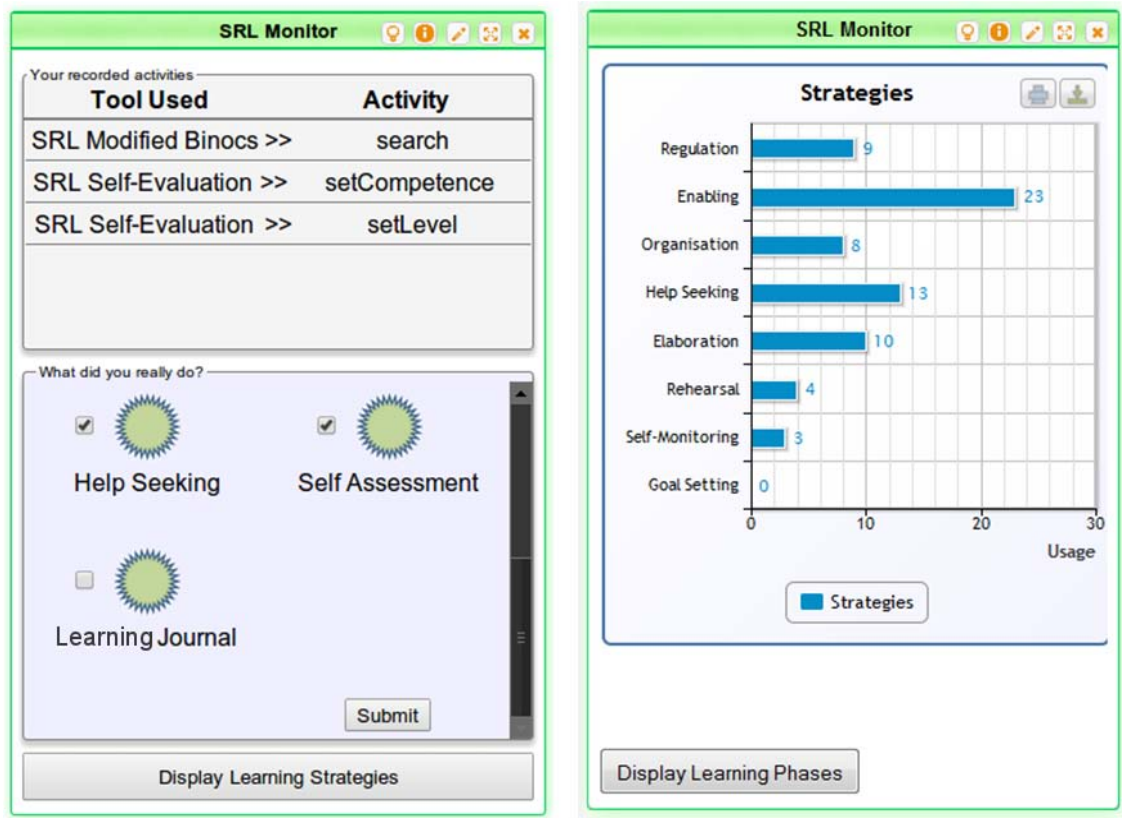

Figure 3.5: The SRL Monitor Widget (two different views).

The second modification of the *Text Reader* bundle consists of only two widgets, the *Text Reader* widget and the *Mashup Recommender* widget (Nussbaumer et al., 2012a). The latter one helps learners to add further widgets, thus customising the bundle. The widget is based on our template approach where mashup designs are modelled on a conceptual level. A template consists of learning activities and learning strategies suitable for a specific or general purpose. In our example (Figure 3.6) we created a template consisting of the activities *collaboration*, *organisation*, and *information seeking*. A learner, who clicks on such an activity, gets respective recommendations for widgets. In our case, widgets for the *organisation* activity are selected and listed in the lower part of the *Mashup Recommender* widget. The learner added the *Share Your Experience* widget, as indicated in the screenshot. This widget allows searching Wikipedia and adding relevant concepts to a personal list that can be shared with peers.

This recommendation approach is based on the ROLE ontology. Each activity is related to certain widgets. By clicking on an activity in the template, the respective widgets are searched in the *Widget Store* and displayed as a search result. Teachers can create templates and learners can make use of these templates. The learners get support to apply a set of particular learning strategies and techniques. This approach is designed to stimulate learners to apply meta-cognitive activities.

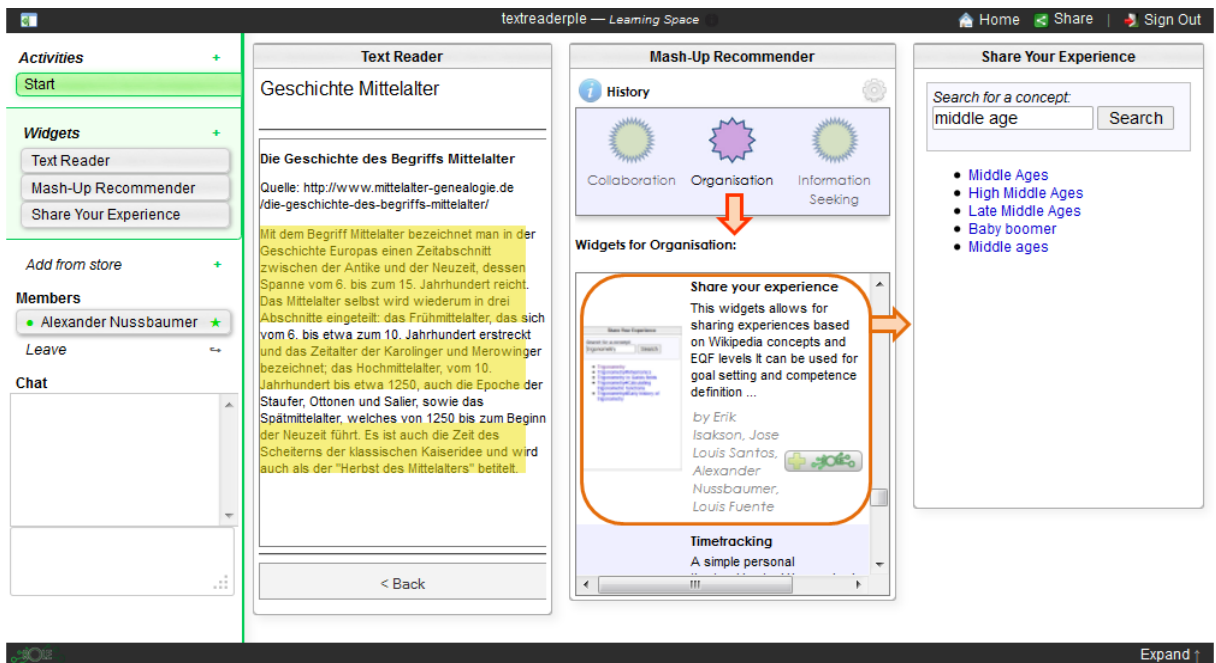

Figure 3.6: ROLE Space with the Text Reader and Mashup Recommender widget bundle in the main area. Additionally the *Share Your Experience* widget is recommended and has been added.

## 3.4 Usage Scenario 2: Collaborative Learning

With the next bundle, the Quadratic Functions Playground bundle (see Figure 3.7), we illustrate the collaborative learning facilities of the ROLE framework. To this end, we show the widgets themselves as well as the surrounding widget space interface, including a list of (online) members and a chat room (left-most frame in the figure). The bundle consists of four widgets for self-regulated learners to reach their goal of understanding quadratic functions. The *Question and Answer* widget enables space members to mutually ask questions, provide answers and to rate the quality of both. The *To-Learn List* widget is used to plan and trace progress for various learning activities. The *Function Plotter* widget is used to interactively vary quadratic function parameters and to study respective graph plots. The *IWC Paint* widget is a collaborative painting canvas used for a function guessing game, where one space member sketches a function graph and the other members have to guess the function drawn.

Figure 3.7: The Quadratic Function Playground Bundle

*Scenario Description: In his high school course, maths teacher Mr. Gauss recently introduced the new topic of quadratic functions. His students Maren and Dominik decided to build a custom PLE on quadratic functions using the school's ROLE platform installation. On his home computer, Dominik points his browser to the ROLE platform and creates a ROLE space "quadratic-functions". Then, he shares the space with Maren by sending her the space's URL via email. Maren receives Dominik's mail and joins the space. Being self-regulated learners, Maren and Dominik decide to use a to-do-list to organize their work. Thus, Dominik navigates to the widget store integrated with the ROLE platform, searches for "to do", and adds the To-Learn-List widget to the PLE. Additionally, for collecting questions and answers during learning sessions, Dominik adds a Question and Answer widget. Maren and Dominik decide to focus on quadratic function graphs first. Thus, they add a corresponding activity item to their to-learn-list. Maren and Dominik plan to support their learning activity with suitable tools. For studying function graphs, they first need an interactive function plotter. Furthermore, they want to test each other with a drawing game, where one of them draws a quadratic function graph, and the other has to guess function parameters. In the ROLE Widget Store, they find a function plotter widget and a collaborative painting widget and add them to their PLE. In a next step, both of them use the Function Plotter to study the graphs of different quadratic functions on their own. After a while, Dominik initiates the graph painting game by sketching a coordinate system and poking Maren via chat. Maren paints a function graph and asks Dominik to guess the function. After*

*studying function graphs and playing their game, they feel confident and thus check the item "study function graphs" on their To-Learn-List. They add some questions to the Question & Answer widget for repetition for the upcoming classwork. Happy with the results of their learning session, Maren and Dominik share their experience with their classmates. After positive feedback from other self-regulated learning classmates, they decide to share their widget bundle in the ROLE Widget Store, such that other maths learners can now benefit from their good practice.*

Starting from the psycho-pedagogical approach, we have demonstrated the theoretical soundness of SRL concepts, its formal and technical realization in the ROLE framework, and its usage in two scenarios. The ROLE approach is the first comprehensive conceptualization of SRL supported by complete technical infrastructure. This infrastructure is introduced in the next session.

# 4 ROLE Technical Infrastructure

This section describes our technical infrastructure and reference implementation needed for enabling SRL with PLE. The system architecture supporting our approach of SRL-enabled PLE consists of five major and interrelated parts: *ROLE Platform*, *ROLE Widget Store*, *ROLE Recommender*, *ROLE SDK* and *ROLE Sandbox* (Figure 4.1).

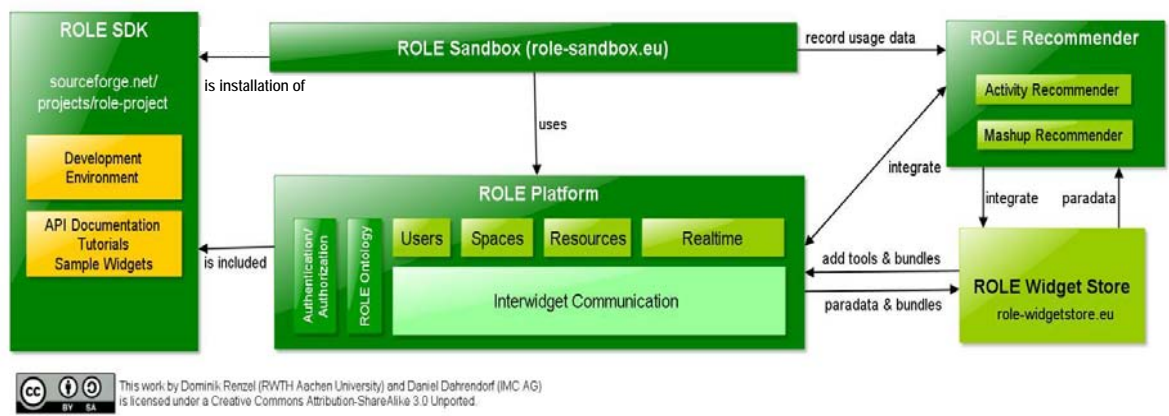

Figure 4.1: Simplified overview on the ROLE system architecture

The *ROLE Platform* is the central component, comprising all different kinds of core and enabling services that support the management and operation of PLEs. *Authentication and Authorization* is realized with a combination of standard OpenID and OAuth mechanisms, but also supports other mechanisms such as LDAP or CAS. *Users* and *Spaces* are dedicated to managing users and widget spaces as well as the respective data stored for these entities, e.g. transcripts of learner activity, learner-created content for users; members, contained widgets, shared data for spaces. Both services are based on a generic service for managing Linked Data resources, thus providing URLs and semantic descriptions for each resource instance compliant with the *ROLE Ontology*. The ontology is implemented in RDF/OWL format and is widely reusing properties from the Dublin Core terms vocabulary. Both services *Users* and *Spaces* integrate with the *Realtime* service, managing different facilities for real-time communication and collaboration among members of a widget space. These means include space-centred chat rooms used for conversations among remote learners and publish/subscribe channels for supporting the communication between widgets in a space, thus enabling the orchestration of single widgets to full mash-up PLEs with the *Interwidget Communication* Application Programming Interface (API). The ROLE Platform supports two-way integration with the *ROLE Widget Store* and a set of *ROLE Recommenders*.

The *ROLE Widget Store* (Dahrendorf et al., 2010) is a repository of learning widgets and widget bundles, enriched with ROLE Ontology-compliant widget metadata, searchable via a SPARQL endpoint. Integration with the ROLE Platform consists of adding single widgets or widget bundles to spaces, sharing best-practice widget combinations as new widget bundles to the store and making widget paradata (i.e. administrative data such as how often a widget was added to a space) collected by the ROLE Platform available to the ROLE Widget Store, e.g. to provide store visitors with information on popularity of resources. Since ROLE Widget Store and ROLE Platform communicate via open specifications, both sides can be easily exchanged.

The *ROLE Recommender* part consists of a set of recommender systems, including the SRL recommendation tools explained in Section 3.2. Many recommender technologies are possible for PLE. Some of them may rely on theoretical learning models. Some of them may rely on observation data only. However, the majority uses both theoretical models and observation data. As such, widget and widget bundle paradata is also taken into account by ROLE Recommenders and accessed from the ROLE Widget Store. Since the principal work of the recommenders is explained in the previous section, here we only mention that the recommenders are realized as widgets. From the perspective of the ROLE Platform, only functionalities assigned to tools are known, but not the pedagogical information. This is also the reason why the recommenders are not part of the ROLE Platform, although they are based on pedagogical information contained in the ROLE ontology.

The ROLE Platform is included in the *ROLE Software Development Kit (SDK)*, which is freely available on the Open Source software hosting platform SourceForge (http://sourceforge.net/projects/role-project). The ROLE SDK also contains all necessary tools, libraries, documentation, tutorials and samples to develop new learning widgets as well as a development environment in the form of a Web application server. This server hosts a frontend for managing, designing, and working with PLEs, in particular to test newly developed learning widgets. As such, it comprises all necessary means and information for everyone to set up ROLE servers in their own infrastructure, e.g. in school or university networks.

The *ROLE Sandbox* (http://role-sandbox.eu/) is a full-featured installation of the ROLE SDK, publicly available on the Web. From the beginning, it served multiple purposes beyond merely serving as a PLE/ROLE platform for learners. We know that the ROLE Sandbox has been used not only for SRL, but also for development and user evaluation of new learning widgets and for scientific or developer workshops. The ROLE Sandbox furthermore serves as a suitable learning analytics research platform (Renzel & Klamma, 2013), whereby different kinds of usage data collected with its help provide the basis for our evaluation results in Section 5.

In the following, we briefly describe the appearance of and the interaction with the ROLE Web User Interface on the example of the ROLE Sandbox. When a learner accesses the ROLE Sandbox she is asked to enter a space name. In a ROLE space, learners are free to orchestrate widgets for their learning activities. The interface includes integration with the ROLE Widget Store such that learners can directly populate their PLE with widgets or widget bundles from this repository. Since this was one of the main ideas of earlier Web widgets environments, e.g. iGoogle as well as of early PLEs, it was also the main motivation for the support of SRL activities by recommendation techniques exemplified in ROLE recommender widgets. As in the ROLE Sandbox widgets can be added arbitrarily to any ROLE space. Also other learners can subscribe themselves to existing ROLE spaces, so that any ROLE space can be turned into a collaborative learning space.

Figure 3.7 (Section 3.3) shows the full user interface for an exemplary ROLE space, consisting of three main areas. The top bar features functionalities to hide and show the left area, access to configuration elements of the space, an exit to other spaces in the ROLE Sandbox, a space sharing tool, and logout. The sharing tool allows for publishing the URL of the current ROLE space on typical social networking sites like Twitter, Google+ and Facebook or to embed the complete ROLE space UI into arbitrary other websites.

The left frame provides means for space management and awareness. The whole space can be subdivided in *Activities*, i.e. different widget constellations within the same space, dedicated to different learning activities. For each activity, a list of all widgets is displayed. Since in many situations not all widgets can be displayed at once, this is a convenient way to navigate to specific widgets. In extensions of the ROLE Sandbox, the widgets can be distributed to other devices of the learner with drag & drop within this area (Kovachev et al., 2013). Also, a button to add widgets is available, which connects to the ROLE Widget Store and allows the widget code to be imported into the ROLE space. Moreover, a list of all space members is displayed together with the information about who is the space's owner (indicated by a star symbol), who is currently online (indicated with green/grey dots) and the option to join or leave a space respectively. Last but not least, the left frame provides a chat window to support real-time communication between all members of the space.

The main area of the ROLE space contains the learning widgets of the currently selected activity. Widgets can be freely arranged and configured in size by the learners. The arrangements are stored and retrieved after a renewed login into the ROLE space. Each widget can communicate with all other widgets in the same space.

The *ROLE Requirements Bazaar* (Klamma et al. 2011, Renzel et al. 2013) is a further development to overcome typical communication problems between developer and user communities, in particular in the SRL area, where the numbers of learners may be outperforming the number of available developers by orders of magnitude. In such a situation, a clear prioritization of needs from the perspective of both learners and developers is very important to realize the most requested features for new or existing learning widgets in the face of limited resources. Therefore, every widget is connected to the Requirements Bazaar (http://requirements-bazaar.org/), an open Web-based platform with Web 2.0 features for voting on ideas or requirements. The idea here is to relate the popularity of learning widgets from implicit measures like downloads with the explicit feedback by rating or voting in the learning communities which are revealed by social network analysis (Wasserman and Faust, 1994).

The purpose of the technical infrastructure is to enable the creation of PLEs in arbitrary learning settings. For this reason it is available as open source software. Parts of it have been already adopted by other open source communities, like Strophe.js. This infrastructure was used for our evaluation studies that are reported in the next section.

# 5 Evaluation Methodology and Results

In this section we introduce our general evaluation approach. In order to demonstrate the acceptance and effectiveness of the ROLE framework and system,, three different kinds of evaluation are presented here: a specific lab study focusing on a short-term learning outcome, a summary of the evaluations in our test beds , and an analysis of usage data collected in the public ROLE Sandbox.. Afterwards we discuss the obtained results and outline limitations.

The goal of the evaluation is to get insights into the challenges of the ROLE framework and its implementation. For our distributed infrastructure and the diverse test beds we adjusted the postulates of the Layered Evaluation Framework for adaptive systems (Paramythis et al., 2010). We applied a multi-method evaluation approach, meaning that different evaluation instruments and methods were deployed. This approach allows a rigorous evaluation of the effectiveness of improving SRL competences integrating short-term and long-term observations.

An overview of the different evaluation tasks is given in Table 5.1. In the first short-term evaluation study we measured the increase of the knowledge level, usability, and workload. This study was user-centred and task-oriented by letting students learn a specific topic with a ROLE space and testing later their acquired knowledge and perception of the technology. The second mid-term evaluation reports the findings of various test beds in informal and formal SRL settings. These findings include reports of teachers and tutors who made use of ROLE technology in their courses. The third long-term evaluation study was centred around the analysis of usage data collected in the public ROLE Sandbox, with a focus on the usage and adoption of the SRL widgets. In the course of two years, the usage of the platform was recorded and analysed considering SRL aspects.

| What is evaluated | Methodology | Who is evaluated | Time frame |
|---|---|---|---|
| Learning Outcome, Usability, Workload | learning task, questionnaires on usability and workload, knowledge tests | 33 students participating at a study | short term |
| Adoption of SRL | reports from teachers or tutors doing learning courses | Approx. 500 users of ROLE test beds | mid term |
| Usage of ROLE platform and adoption of SRL | ROLE Sandbox usage data analysis | Approx. 1400 users of ROLE Sandbox | long term |

Table 5.1: Overview of the evaluation methodologies and related results.

## 5.1 Short Term Evaluation of Learning Outcomes

We carried out this lab study with university students. The goal was to find out how effective and usable a predefined widget bundle was as well as if the concept of creating and modifying widget bundles was accepted by the learners. To this end the SRL Text Reader bundle introduced in Section 3 was used in different settings.

**Preparation and Pre-tests**

In the beginning phase of the study the preparation of the technical settings took place. Two environments were prepared – the static *SRL Text Reader* bundle consisting of *Text Reader*, *Self Evaluation*, *Self-Reflection* and media recommendation (*Modified Binocs*) widgets (Fig. 3.3) as well as the flexible SRL Text Reader bundle consisting of the *Text Reader* widget and the *Mashup Recommender* widget (Fig. 3.5). As described in Section 3, the Text Reader widget displays pieces of text according to a predefined domain model. For the evaluation we used a specific knowledge domain, unknown by the students. It was History of the Merovingian dynasty who ruled over France in the Early Middle Ages.

The second phase consisted of preparing questionnaires for usability and knowledge tests. Here, we employed the System Usability Scale (SUS), generally consisting of ten items (Brooke, 1996). For this case study, our SUS questionnaire contained answer options that consisted of a five-point Likert scale ranging from "strongly disagree" to "strongly agree". The generated raw data from the survey was then computed to an overall SUS score, as suggested by Brooke.

The NASA task load questionnaire (NASA-TLX) is a measurement instrument which assesses the workload of individuals using six subscales with a score range of 0-100 where higher values indicate a higher workload (Hart, 2006). The contained subscales are *mental demand*, *physical demand*, *temporal demand*, *performance*, *effort*, and *frustration level*. An overall workload score is calculated based on the mean of all item contributions.

In order to assess any learning gains by students during a fixed period of study, we administered two knowledge tests related to the subject of the Merovingian dynasty. One was deployed as a pre-test at the beginning of the learning period with the prepared PLE. The other test was a post-test right after the learning period with the PLE. The tests required students to learn about the Merovingian and the Middle Ages by reading related texts. Each text contained 17 multiple choice questions.

The third phase included group formation according to SRL competences. In summer 2012 a newsletter email was sent to approximately 22.000 students of the University of Graz to participate in the study. A cinema voucher was promised as an incentive for the participation considering the fact that these students had to be physically present in our office. The newsletter also contained a link to an online questionnaire relating to SRL competences (QSRL; Fill Giordano et al, 2010). On a scale from 0-100 the participants had to answer 18 questions, for example how well they prepare their learning plan in terms of time management or how much they think about their learning outcomes after a learning episode. The outcome of this questionnaire includes an overall assessment of the learner's SRL competences by herself. 153 students indicated that they were interested in the study. The SRL data was analysed and the respondent student cohort was split into a high and low level SRL group by dividing the students according to the median of their competence levels.

Out of the 153 interested students 33 actually participated in the study. According to their SRL competence levels, 15 students were below the median of the SRL competence levels. 18 students were above the median. We created four groups (outlined in Fig. 5.1). Two groups (SRL low and high) were asked to use the static *SRL Text Reader* bundle. The other two groups (SRL low and high) were asked to use the flexible SRL *Text Reader* bundle.

| Groups | Static SRL Text Reader bundle | Flexible SRL Text Reader bundle | Sum |
|---|---|---|---|
| SRL level low | 7 | 8 | 15 |
| SRL level high | 9 | 9 | 18 |
| Sum | 16 | 17 | |

Figure 5.1: Distribution of the participants regarding PLE environment and SRL level.

The fourth phase included the analysis of demographic data of the 33 participants. We prepared a questionnaire to get a profile of the learners. A series of survey questions was constructed that determines the age, highest educational degree, profession, country of origin, experience with widgets or apps, hours of using the computer per week, and experience with learning software. The analysis of the collected data shows that the 33 participants were on average 24 years old, had heterogeneous educational background (psychology, economics, art, chemistry, geography, etc.), and were from Austria, except two, who came from Italy and Croatia, respectively. There were 20 females and 13 males. In addition, the participants also reported that they used a computer more than 10 hours a week (60.61%). 33.30% of them declared that they rarely work with apps or widgets. 81.80% stated that they rarely learn how to use new software. Also the majority of participants indicated that they had no previous experience with the widget environment iGoogle before the evaluation took place (75.80%).

Finally, the 33 participants had to individually perform the learning tasks with the prepared PLE setting. This part of the lab study then started with the pre-knowledge test. Then the students were asked to learn with either the static or the flexible SRL *Text Reader bundle*. They were introduced to these PLEs by the instructor who informed that they should also prepare for a post-test. Then participants had to use the PLE for approximately one hour and learn the contents of the prepared texts on the Merovingian dynasty. After the learning period we asked participants to answer the post-knowledge test and to complete short questionnaires on usability.

**Results**

According to the four experimental groups (SRL low and high, static and flexible SRL Text Reader bundle) and two knowledge tests (pre- and post-test), eight average knowledge values were calculated (see Figure 5.2). We observed that the groups had different levels on the pre-tests. One low level SRL group had the highest level and the other low level SRL group had the lowest level. The post-test revealed that all groups came to almost the same level, independent of the SRL level and the PLE group (which seems to be a ceiling effect). Most importantly, for all groups higher scores in the post-test were observed compared to the pre-test. Finally, a t-test for repeated measures showed that the scores of the post-test (M=14.64, 1.52) were significantly higher than for the pre-test (M=7.12, SD=2.52) for all participants ($t_{33}$=16.78, p<.001).

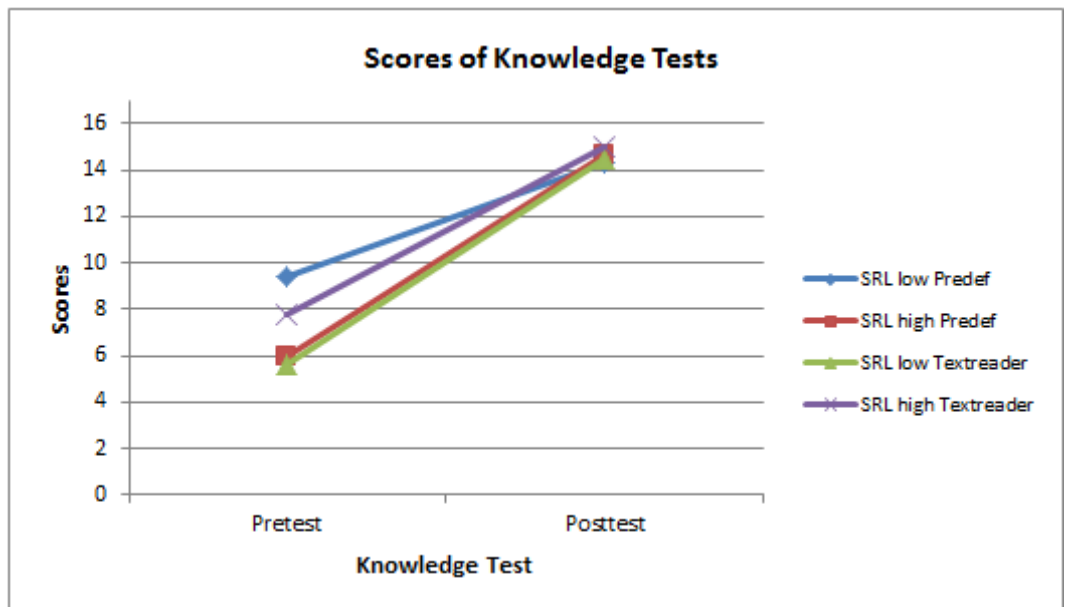

Figure 5.2: Scores of the knowledge tests (pre- and post-test) for the four experimental groups.

Looking more closely at the usability related aspect of the analysis it can be noted that a usability score of 71.27 (SD=17.50) was obtained. Given a possible score range of 0-100, with higher values indicating higher usability, this result indicates that the usability is between medium and good. However, a high value of the standard deviation shows that there were also learners who gave a rather poor rating. There were no differences observed for both experimental groups. Therefore, only the mean scores of all participants are presented in this description. The individual values of the SUS subscales are depicted in Figure 5.3 in a range between 0 and 4 as it was questioned received on a five-point Likert scale. There were no reports on problems on the mixed-language interface (user interface in English and content in German). However, this is no surprise, because the average Austrian student is used to English language in computer environments.

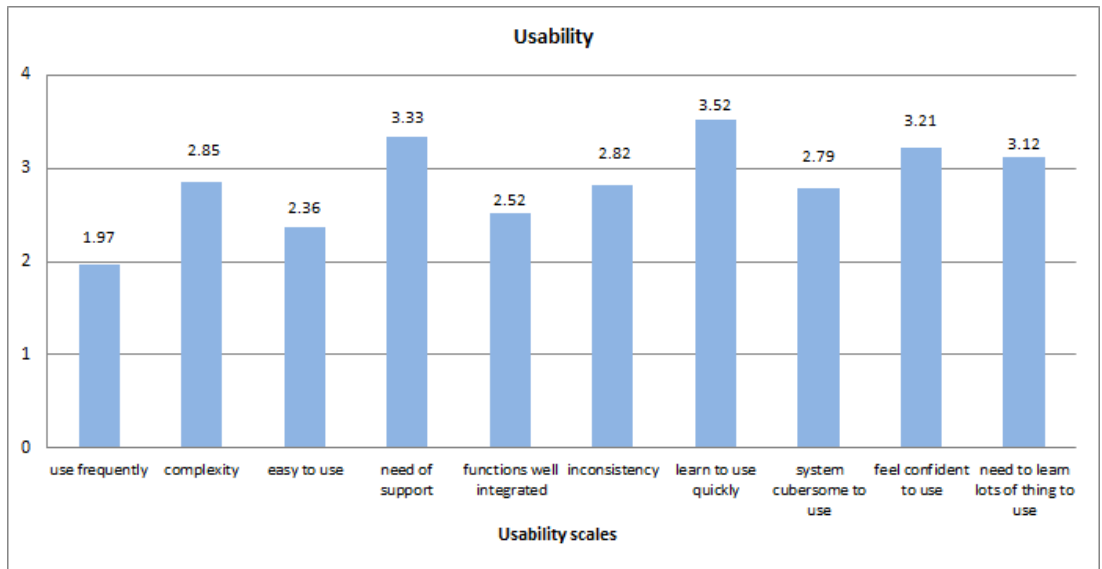

Figure 5.3: Items of the SUS usability test. These are values as received on a five-point Likert scale from 0-4. In this diagram negative questions have been inverted so that higher values are better for all items. Mapping these values on a scale from 0-100 the average value is 71.27.

The results of the task load questionnaire indicate that the workload of learning with the PLE was perceived as low by the participants. Participants reported the mental demand and effort as the highest, but still moderate workloads. These subscales refer to the mental and effort resources that had to be mobilized to accomplish the task. Consequently, we would suggest that the result for effort can be relegated to the mental category similar to the demand. In this context the concept of mental effort can be described as the voluntary mobilisation of mental energy or resources of attention towards a task (Kahneman, 1973). If task difficulty increases then engagement increases, too. The analysis of our results suggests that the difficulty level was perceived as medium. The lowest scores were recorded for physical demand, frustration and temporal demand. It is not surprising that the physical demand was rated low. It is interesting to note that the learners did not appear to become frustrated during the learning session, which is the basis for a successful and optimal learning experience. Finally, it can be reported that no statistical differences were found for the two experimental groups concerning workload. Therefore Fig 5.4 displays the mean for all participants (N=33).

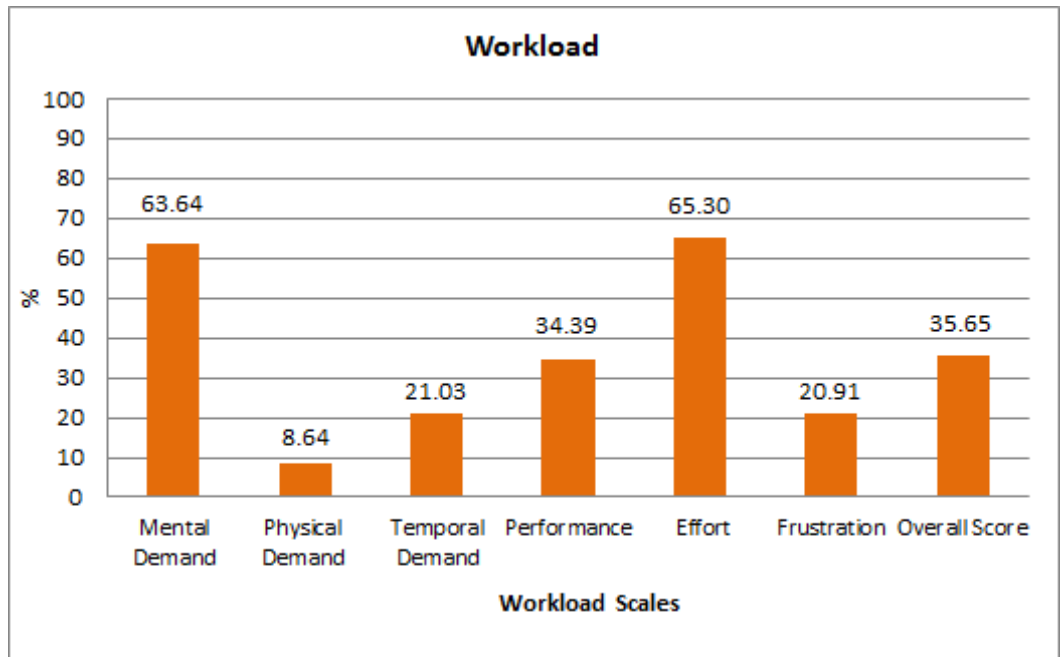

Figure 5.4: Workload score (N=33)

The analyses of log data reveals that the learners of the static *SRL Text Reader* bundle (N=16) performed a relatively high number of SRL events (see Figure 5.5). The events *Tags added* and *Tags removed* are performed in the *Text Reader* widget. The event *Competence* is performed in the *Self-Evaluation* widget when a learner assigns a concept to her own competences. The event *Searched Tags* occurs when a learner requires information on concepts in the *Self-Evaluation* widget and gets a result related to learning objects in the search (*Modified Binocs*) widget. The event *Feedback* occurs when a learner accesses a visualisation in the feedback (*Self-Reflection*) widget. These SRL events are mapped to SRL learning strategies. Adding and removing tags is mapped to *elaboration*, defining competences is mapped to *self-evaluation*, and requesting feedback is mapped to *self-reflection*. The learners of the flexible *SRL Text Reader* bundle (N=17) added and used 95 widgets in total. This activity is mapped to the SRL learning strategy *environment preparation*. This observed behaviour demonstrates that learners performed SRL activities on the cognitive and meta-cognitive level and thus proves that learners were willing to learn self-regulated.

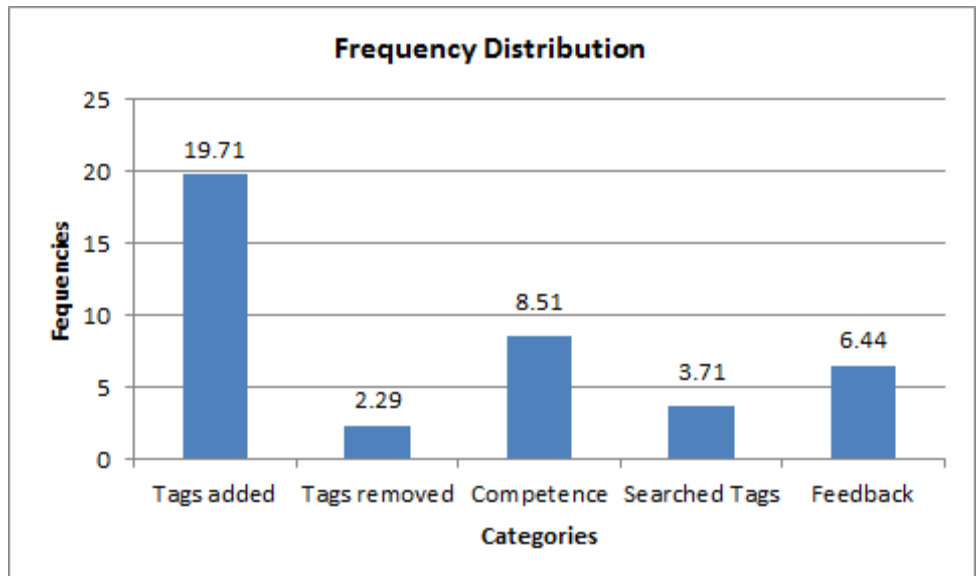

Figure 5.5: Distribution of SRL events for the static SRL Text Reader bundle group. The numbers display the means for N=16.

The major finding of this evaluation study is the demonstration that learners are willing to use widgets bundle and apply SRL strategies and techniques. Though results from the usability and task load questionnaires suggest room for improvement, the learning outcome still significantly increased. However, a limitation of such a short-term lab study is that an increase of applying SRL activities or an increase of SRL competences cannot be measured. Therefore, in the following we focus on mid-term and long-term studies in authentic SRL settings. Consequently, the level of control in these settings is limited, but with appropriate measuring and monitoring techniques still valid observations can be made and analysed.

## 5.2 Mid Term Evaluation of Test Beds

The mid-term evaluation results of the ROLE approach have been achieved in the test beds (reported in Mikroyannidis & Connolly, 2013). All these test beds represent a wide variety of learning contexts and case studies. Their outcomes could help understanding the challenges and opportunities associated with the adoption of PLE in different learning settings. In order to be consistent in the evaluations and to enable their comparison, a common survey was prepared. It aimed to measure the Perceived Usefulness and Ease of Use (PUEU) of PLE. Each test bed could enhance this core part with questions targeting its specific context. The respondents included various types of users, e.g. educators, learners, employees, and trainers. Moreover, a questionnaire on teachers' perception of SRL in the higher education (TPSRL) was devised. In the following we present relevant outcomes of three test beds as well as a summary from all eleven ones.

In the test bed at the RWTH Aachen University in Germany, the concrete setting was a lecture on computer science in mechanical engineering, where about 1600 students attended the classes in 2012. Its practical part, the project laboratory was developed in response to previous student evaluations requesting "hands-on" tasks. In previous iterations of the lab course students could only used the Web Knowledge Map, similar to a semantic wiki. In order to facilitate SRL in a PLE, three ROLE widgets were created and deployed – the Web Knowledge Map (WKM) widget, a chat widget and a history widget. Interviews with the lecturer and three assistants were carried out after the course. The results compared with the physical attendance at lab courses indicated students' positive judgment of PLE, better supported autonomy of students, more efficient explanations from the tutor, improved communication among students, and their more flexible time management. Among the weaknesses, technical and usability issues in the administration of the environment were reported. Several extensions of the PLE have been suggested, including a learning planner, a search engine, and a recommender system for external materials. The teachers also appreciated the chat tool, as this feedback helped them to estimate students' progress and to adjust interventions, also making teaching more satisfying for tutors. So, this evaluation showed that the PLE helped to support SRL and collaborative learning. The ROLE widgets enabled the students to mutually support each other. Nevertheless, the change of formal settings has to be performed carefully, by focusing on activities for which the SRL and PLE benefits are significant enough.

The test bed at the SJTU University in Shanghai (China) followed a traditional pattern with a teacher-centric focus, using a "broadcast" model, where students usually watch lectures rather passively. For this test bed, ROLE offered enhanced opportunities for interaction between learners and teachers. In the language learning setting (e.g. French, German, business English) it included widgets enabling students to practice their pronunciation by recording themselves and comparing their speech with the "original". The evaluated courses used a PLE based on the ROLE Moodle Widget Space, i.e. a plug-in allowing the usage of widgets in Moodle. In this environment, teachers set up several ROLE-enabled online learning and teaching spaces, including domain-specific content and widgets. Students were asked to complete a survey after they finished their course of study, but it remained voluntary. 20 out of 150 students from the French and German courses participated in the study. Regarding their knowledge of e-learning, 65% felt that it was quite good and 35% considered it as high and better. According to the results, 65% of students used the PLE, 20% did not use it, and 15% did not know what the term PLE meant, despite their self-stated fluency in e-learning. Students also reported a positive experience of using the PLE. Based on a Likert scale (1 = strongly disagree, 5 = strongly agree), all but two students agreed or strongly agreed with statements on PLE usefulness, effectiveness for their work, helpfulness, and ease for use. With the negatively formulated statements about frustration from PLE usage and high mental effort requirements 7 students agreed or strongly agreed, while 13 were neutral or disagreed. The majority of students (15 or more) agreed that they would like to modify their PLE by adding or replacing widgets. Although the students thought that the prepared PLE/Moodle space was well designed, the actual usage logs indicate that only a few ROLE widgets were used by them. Thus, the survey results suggest potential interest rather than active ROLE widget usage. Three teachers provided semi-structured interviews and reported that ROLE was extremely useful as it allowed learners to access materials in a more flexible manner, enabled them to self-assess their skills, and provided enhanced access to the collected metadata for teachers and students. Two areas of improvement have been suggested in the interviews, namely the need for more subject-specific widgets and better usability of these technologies. So despite some technical difficulties experienced at SJTU, it turned out that a different cultural background influences SRL attitude and the use of PLE technology.

The test bed in the training centre of the German engineering company FESTO mainly addressed the technical challenge: integration of a PLE with the existing Learning Management System (called Clix) into a "Personal Learning Management System" (PLMS). The aim was to enable an advanced search and simplified access to relevant learning materials, to support the planning and the reflection of the learning activities, as well as to raise motivation and promotion of SRL and different forms of cooperative learning. In the evaluation a focus group of 26 participants were asked to complete a survey. The majority of the respondents approved the look-and-feel and the usability of the used ROLE widgets. However, a big challenge was to promote this new approach on a large scale. In this industrial setting also other issues appear, including openness versus data security, different interests of the company and the individuals, as well as the implementation strategy (revolution versus evolution). Nevertheless, it was possible to deploy an integrated PLMS with predefined learning spaces adjusted to the SRL competences of the learners.

So what are the main lessons learned from a wide variety of learning contexts and case studies in the ROLE test beds? First of all, for most of the test bed populations SRL and PLE were new concepts. Therefore, it was necessary to raise awareness on them and explain them properly. For this purpose, ROLE prepared various introductory and guidance learning materials in different formats. Early adopters among teachers and trainers needed special assistance in order to foster a culture where the community is willing to engage with new innovative technology. These users required accessible and easy to use widgets that could be used effectively and efficiently. They often needed tailored tools for specific subjects and contexts, as well as widgets that harvested information from suitable repositories. Moreover, effective evaluation and feedback mechanisms were crucial. For this purpose, ROLE adopted a Social Requirements Engineering approach (Renzel et al. 2013), allowing users and developers to communicate and collaborate within a social networking environment. The ROLE experience has shown that this approach to support SRL with PLE technology provides flexibility for various settings and enables integration with various learning platforms, which can be considered as a very valuable outcome with the potential to improve the educational process in general.

## 5.3 Long Term ROLE Sandbox SRL Analysis

In this section we present the results of an analysis of the ROLE Sandbox usage data. Here we followed the approach from Renzel and Klamma (2013) with special regard to SRL tool support and usage found in the data. The data set used in this work comprises usage data from roughly 2 years of ROLE Sandbox usage from March 3, 2012 – March 20, 2014. A single data entry contains information such as origin IP address (who), timestamp of request (when), requested resource (what), user agent (which browser), status code (success vs. failure), etc. Regarding IP addresses we cleaned the dataset from entries caused by automated agents such as search engine crawlers or malicious bots to restrict analysis to human-operated agents only. Furthermore, we filtered out all requests coming from IP addresses affiliated with ROLE partner institutions, thus only taking into considerations ROLE Sandbox use from external IP addresses. For the remaining IP addresses, we employed an IP geo-location service to enrich the data set with geospatial information from where the particular request came. Furthermore, we restricted our analysis to only those entries that relate to the usage of the ROLE Sandbox's RESTful APIs. Requests to static content were not part of the analysis, because they give no extra information about carried out activity. ROLE Sandbox API requests enable the extraction of operations such as *create a space, join/leave a space, add/remove a widget to/from a space*, etc. Additional widget meta-data related to the SRL phases, in particular the set of categories defined in the ROLE Ontology (Kiefel, 2013) was retrieved from the ROLE Widget Store (Nussbaumer et al., 2012, Dahrendorf et al. 2012) via its SPARQL endpoint and merged with the dataset.

In the 2 year period of ROLE Sandbox operation until now we collected 2.52 Mio API requests in total. 0.68 Mio (26.98 %) of these requests originated from over 4800 distinct IP addresses from 705 cities in 89 countries.

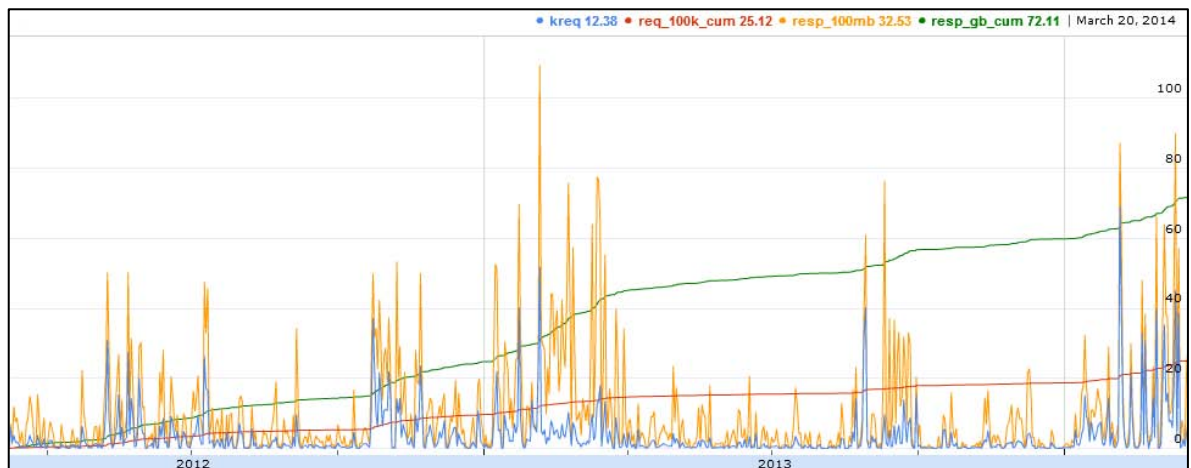

Figure 5.6: Temporal distribution of ROLE Sandbox usage.

Figure 5.6 provides an overview of the temporal distribution of ROLE Sandbox usage by aggregating and cumulating daily statistics: kilo-requests served (kreq; blue graph); cumulative hundreds of kilo-requests served (req_100k_cum; red graph), hundreds of MB served (resp_100mb; yellow graph), and cumulative gigabytes served (resp_gb_cum; green graph). Figure 5.7 shows the geospatial distribution of ROLE Sandbox usage, limited to only those external IPs having accessed the ROLE Sandbox API more than 10 times. The dots on the map aggregate different numbers of distinct IP addresses resolved to the same geo-coordinates, encoded with different dot symbols and colors. High numbers of distinct IP addresses with the same geo-location hint to ROLE Sandbox usage from a larger organization. Low numbers indicate the use by individuals. The ROLE Sandbox was used by both individuals (single IPs) and institutions of varying size (multiple IPs at same location) and varying intensity. Usage is mainly concentrated in European countries, but also includes bigger institutions in both China and the US.

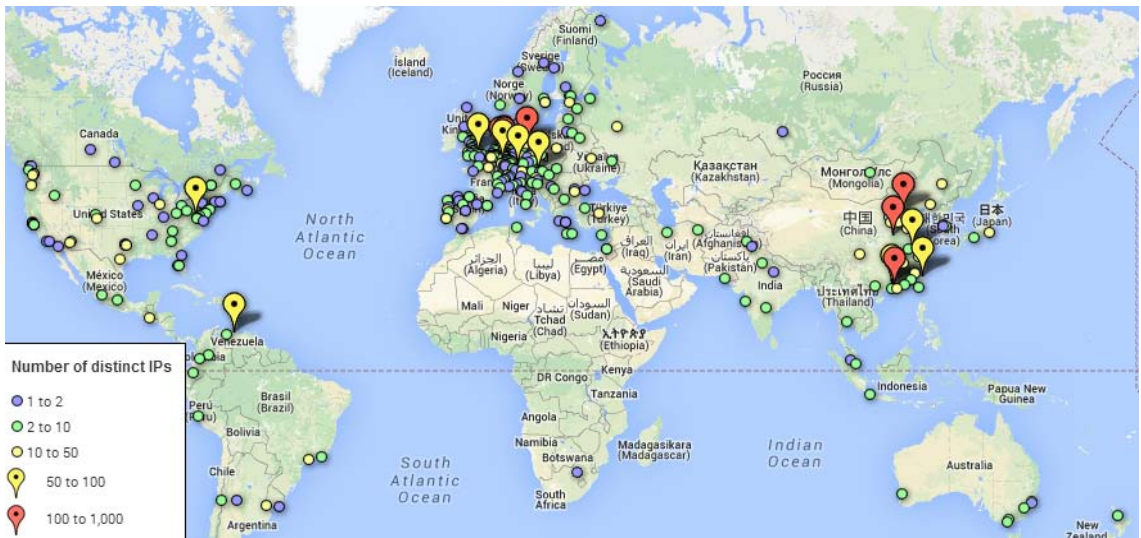

Figure 5.7: Geospatial distribution of ROLE Sandbox usage (distinct IP addresses per coordinate set).

In total, our external users created 187 spaces, where 53% of them can be considered active spaces, i.e. spaces which were loaded frequently. The other 47% were spaces where users tested ROLE for short time periods. About 1400 different external users actively used the ROLE Sandbox in different ways: 118 (8.3%) users created new spaces. 293 (20.6%) users joined spaces created by other users. 353 (24.8%) users added widgets to spaces. 1320 (93.1%) users re-opened spaces with widgets. These statistics confirm the usual observation in social software systems that only a small number of users become active in terms of designing PLE on their own, while most other users only benefit from PLE previously created by others. By definition, one part of SRL includes the design of the user's PLE as well as its refinement after repeated self-reflection and evaluation. Thus, we can assume that only a small number of ROLE Sandbox users are self-regulated learners in that sense.

Furthermore, we analyzed the widgets used in the ROLE Sandbox with regard to SRL. In total, external users employed 379 different widgets in their learning spaces. In order to support SRL, we designed and implemented a set of 15 SRL widgets for different purposes such as recommendation, self-assessment, self-monitoring, self-evaluation, and SRL-style work. In our data we located 11 (2.9%) SRL-related widgets used by external users.

All widgets imported into the ROLE Sandbox from the ROLE Widget store were assigned to one or more of the seven categories *"Search & Get Recommendation", "Plan & Organize"*, *"Communicate & Collaborate"*, *"Create & Modify"*, *"Train & Test"*, *"Explore & View Content"* and *"Reflect & Evaluate".* They map to the four phases of the SRL model presented earlier. Fig 5.8 provides an overview, which widgets are explicitly SRL-related (red nodes) or not (grey nodes), and to which categories (blue nodes) they were assigned. For widget nodes, node size encodes how often the particular widget was added to a space. For category nodes, node size encodes how many widgets are assigned to the particular category. The large set of nodes in the bottom left, where none of the nodes is connected to a category node indicates that most widgets were not assigned to any category, including the most influential SRL widgets (unconnected red nodes). Most of the assigned widgets are only associated with exactly one category, thus indicating their clear purpose for one of the SRL learning phases described earlier. For example, the *Share Your Experience* Widget (Section 3.3, Figure 3.5) is assigned to the category *Reflect & Evaluate*, while the *SRL Activity Recommender* Widget is not assigned to any category.

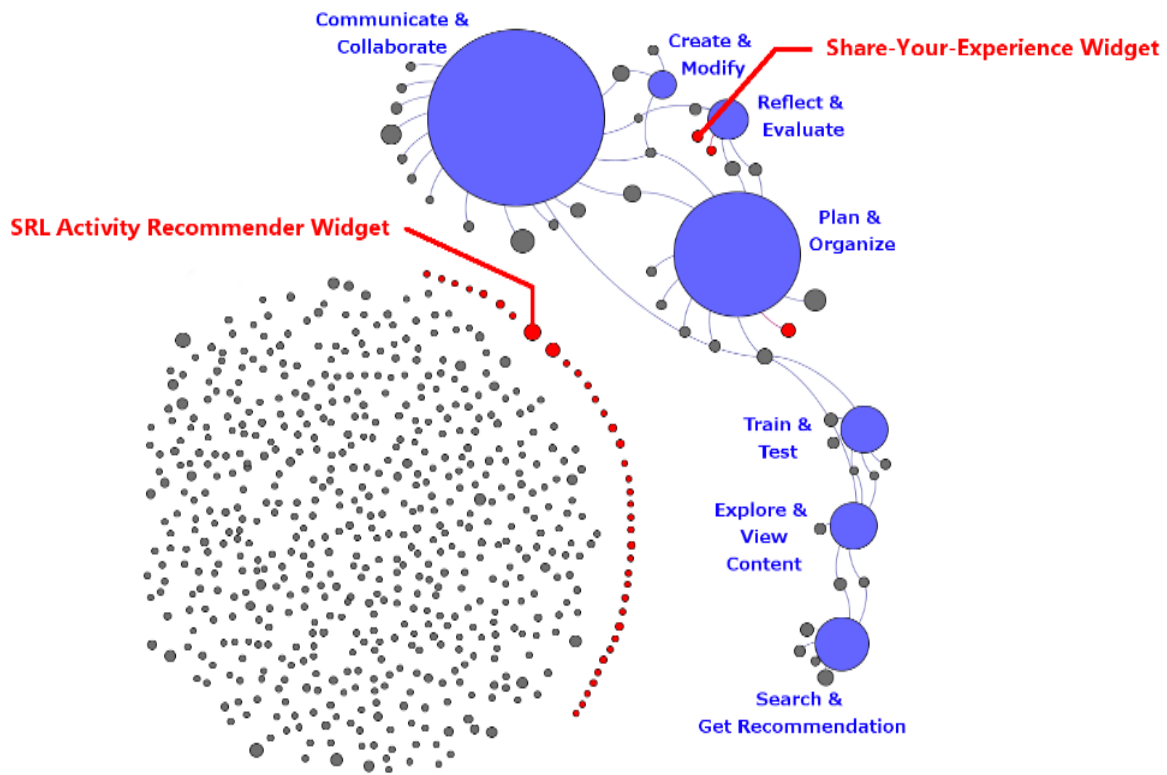

Figure 5.8: SRL and non-SRL widgets and their assignments to SRL categories.

Furthermore, we analyzed our data from the space perspective with the goal to identify SRL activities in spaces. In total, we found 143 SRL-enabled spaces, i.e. spaces where SRL widgets were added and loaded. 31.47% of all SRL-enabled spaces can be considered active, i.e. widgets were added and loaded frequently later on. External users were involved in 40% of these spaces. In 15.38% of all SRL-enabled spaces, SRL widgets were added by both externals and partners. In 29.59% of all SRL-enabled spaces, SRL widgets were loaded by both externals and partners. We can thus say that ROLE partners and externals collaborated in both designing and using these spaces and that certain SRL training was given by ROLE partners to externals. Several SRL-related workshops conducted by us explain this finding. In order to find out if external users would employ SRL widgets on their own, we had a closer look to the 11.2% of SRL-enabled spaces designed and used exclusively by external users. Although the number of such spaces is quite low, we found recurrent patterns. The use of different SRL activity recommender widgets helped learners to design PLE for their learning purposes. Furthermore widget usage patterns in these spaces support different phases of our SRL process model. In particular, we examined widget add and load activities in these spaces. Most of them were designed and used by single users. In rare cases, such spaces were collaboratively designed and used by multiple users. In general, we found that SRL-enabled spaces are used quite productively, but only over short time periods, on average 21 days. However, after the use of SRL widgets, users tend to more actively create and design new spaces, targeting specific learning activities. It seems as if spaces are considered as resources that are only relevant for certain focused (short-term) learning activities.

We highlight one learning space, where multiple external users from two different cities in Germany collaboratively designed and used a language learning space over 28 days. Immediately after space creation, the users arbitrarily added and removed two different widgets mainly dedicated to communication and collaboration. Any inferences on particular learning goals pursued in the space were not possible at that point. Three days later, one of the users added an SRL activity recommender widget. On the same day, the same user added a set of widgets for learning a foreign language (vocabulary trainer widget, foreign media search widget, media player), for planning and reflecting on their learning process (*TODO list* widget, *Share-Your-Experience* widget), and for communicating with each other (multi-user chat widget). This widget constellation was then kept for the remaining days.

Finally, we examined widget add operations from the category perspective to receive an overview which categories became most influential in SRL spaces and in spaces in general (Figure 5.9). First, widgets added to spaces belonged to no specific category for SRL spaces (58.8%) in comparison to non-SRL spaces (64.8%). Furthermore, we found differences with respect to the importance of categories for SRL and non-SRL spaces. While non-SRL spaces put the strongest focus on the category *Collaborate & Communicate*, SRL spaces are stronger in support for SRL-relevant categories such as *Plan & Organize* (13.0 % vs. 8.7%) and *Reflect & Evaluate* (4.7% vs. 2.6%). In other categories, differences between SRL and non-SRL spaces seem rather small.

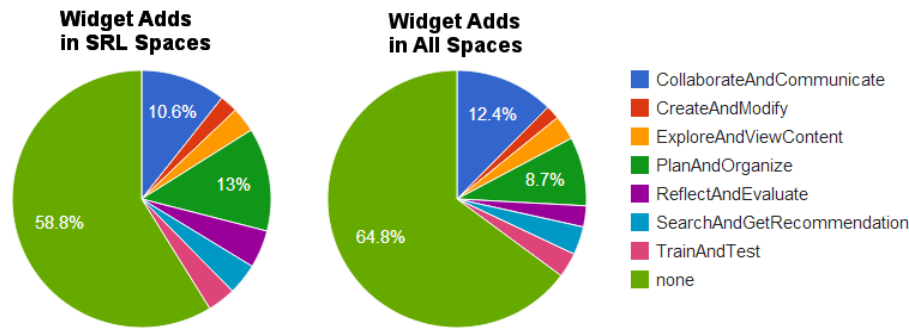

Figure 5.9: Categories of widgets added to SRL spaces vs. all spaces.

As a result of our analysis of the ROLE Sandbox usage dataset we can conclude that SRL widgets supported learners in better organizing and reflecting their learning process and in finding suitable learning resources. It should be noted that our analysis only shows slight evidence of SRL benefits, but cannot prove that SRL took place at all. The data merely captures PLE management operations, and not concrete learning activities. However, we consider PLE management operations as integral part of SRL, in particular in planning and exploration phases, and thus we see this long-term analysis as partial contribution to measuring if SRL has happened in the ROLE Sandbox.

## 5.4 Discussion and Limitations

SRL is a complex competence that must be cultivated over a long time to become effective. Nevertheless, to show its acceptance and effectiveness, we decided to take three different perspectives here – a user study focused on short-term learning, the feedback collected in several various test beds, and long-term SRL data analysis on one of the deployed learning platforms. Together, these views illustrate the potential and drawbacks of this approach.

An important finding was that many students are not familiar with SRL and expect a lot of interventions from their teachers, as done in traditional education. This may have negative consequences later on in their lifelong learning. Unsurprisingly, teachers like working with self-regulated learners. Nevertheless, in the traditional curricula there is not much opportunity for the development of the SRL competences of students. However, students easily understand explanations of the SRL and PLE concepts. But to design a PLE for a particular purpose, even university students rely very much on the teachers' expertise. Perhaps this can be improved over time as they become more familiar with the technical environment, which has to be intuitive and robust in order to attract users. However, it turned out that just providing technology is not sufficient to cause a change in the learning culture. For both instructors and learners, training courses and introductory materials on SRL were as helpful and motivating.

Although the short-term study did not prove that learning effectiveness depended on the level of guidance and freedom in this case, it demonstrated that an increase of the knowledge level could be measured. Usability of the PLE and workload could be improved, but still the learners were willing to use the system and did not got frustrated, but were interested in the further usage of ROLE technology. This study also demonstrated that the conceptual approach as described in Section 3 and implemented in the used widget bundles represents a valid way to support learning by technology. The analysis of the log data showed that learners applied SRL activities, which we regard as effective usage of the PLE.

The long-term analysis of data collected on the ROLE Sandbox confirmed its regular usage and a wide coverage regarding its users, but on the other hand it also showed rather limited use of SRL related PLE in the captured time frame. This revealed a lot of potential in further development of useful SRL supporting widgets as well as in their usage in practice. Realizing this, we have decided to keep this platform open in the future for research purposes, in order to measure long-term aspects of SRL, like engagement and effect. Part of this initiative is the development of innovative tools that would not only support learners and teachers, but also researchers in this case. Taking into account the evaluation tasks already performed, we could demonstrate that the platform itself can be used as a research platform where further studies on SRL can be performed. This opens up a particular opportunity for long-term studies on the effectiveness of SRL.

The ROLE approach provided opportunities to observe long-term authentic behaviour of learners. But, compared to controlled lab studies, the level of control is much less. Additionally, qualitative long-term research methods, like ethnography, can be deployed.

This is already common in disciplines like software engineering. The community of practice approach by Wenger (1989) could be an ideal companion to our quantitative data analysis approach. SRL cannot be just learned as a domain knowledge, but it takes longer time periods and their adoption is highly socially driven (Pressley, 1995). In addition, a valid measurement of SRL competences requires comparable studies with same conditions. However, an evaluation of SRL competence increase on that basis requires more resources than were available. A second limitation was the relatively low number of widgets available, as the Widget Store was relatively new. Ideally, a large number of widgets related to domain knowledge and SRL would be needed to provide more choices and to get a deeper insight into the learners' behaviours with the ROLE platform.

# 6 Conclusions

In this paper we dealt with support of SRL by means of ICT, which is crucial for cultivation of lifelong learning competences. A framework has been developed that allows establishing support strategies for SRL by creating widgets and widget bundles. Moreover, the ROLE platform allows compiling, arranging and using those widgets in a PLE. ROLE technology enables also to find suitable widgets and to easily integrate them in a PLE.

From our perspective, the main contribution of this paper is a framework for integrating SRL support in Responsive Open Learning Environments. The ROLE framework aims to support students to develop SRL competences. A good SRL solution should be personalised, providing a right balance between the learner's freedom and guidance, and motivating the learner. Therefore, we elaborated the operationalisation of the SRL process and implemented a user model for SRL. On that basis, a technical approach including recommendation and monitoring has been realised. Our approach is inspired by the principles of libertarian paternalism, providing enough freedom for the learner, but also suitable nudges to guide her. That means, the information in the learner model is exploited to provide meaningful recommendations and feedback for reflection, rather than automatic adaptation of the learning environment. This may be considered as novel compared to traditional adaptive systems.

Evaluation of the effectiveness and usefulness of our approach has been undertaken in different settings, involving both students and teachers. The overall findings suggest that this is a promising direction for further research and development. However, behavioural changes in this field have also limits and barriers and require long term research. Of course, we also realise certain limitations of a purely technical approach. According to Fogg (2009), triggers fail if motivation and ability is low. In such situations human support can be more effective.

ROLE outcomes will be used consequently in further research contexts. The ROLE platform turned out to be an environment for experimentation and further research on SRL approaches. Creating new widgets and taking up the different evaluation methods can lead to new findings on SRL support. For example, the new EU project Learning Layers (http://learning-layers.eu) intends to take up these ideas. It develops a set of modular and flexible technological layers for supporting workplace practices that unlock peer production and scaffold learning in networks of SMEs. This should bridge the gap between scaling and adaptation to personal needs. By building on recent advances in contextualised learning, these layers provide a meaningful learning context when people interact with other people, as well as digital and physical artefacts for their informal learning, thus making learning faster and more effective. The project builds on mobile learning research and deals with learning in physical work places and practices to support situated, faster and more meaningful learning. The focus of the lifelong learning project BOOST (http://boost-project.eu) is the determination and monitoring of learning goals in small enterprises using the ROLE platform and newly designed widgets.

# Acknowledgements

The work reported has been partially supported by the ROLE project, as part of the Seventh Framework Programme of the European Commission, grant agreement no. 231396.